# The Global Carbon Negative and the "Scope 4" Neutralized Carbon Emissions

Zhu Liu[1]*, Guangqian Wang[2]

[1] Department of Earth System Science, Tsinghua University, Beijing 100084, China

[2] State Key Laboratory of Hydroscience and Engineering, Department of Hydraulic Engineering, Tsinghua University, Beijing, 100084, China

* Corresponding author

E-mail: zhuliu@tsinghua.edu.cn

**Abstract**

Assessing carbon negative and carbon neutrality is critical for mitigating and adapting global climate change. Here we proposed a new framework to account for carbon-negative and carbon-neutral actions by introducing the definition of Carbon Negative (C0), Carbon Neutrality Stock (C1), Carbon Supply (C2) and "carbon-neutral emissions" or "Scope 4 emissions," which refers to the "avoided emission" due to use of non-fossil energy or C1 products. For the first time, we calculated the global neutralized carbon emissions or "Scope 4 emission" by renewable electricity generation, and the results indicating the significant contributions by China, with total neutralized carbon emissions (2.15 Mt C/day) much higher than the U.S. (0.85 Mt C/day) and EU27 & UK (1.25 Mt C/day) together. We show that China contributed to more than 36% of global neutralized $CO_2$ emissions, and such contributions are still increasing. This new framework reflects China's remarkable contributions to the global climate mitigation through the development of carbon neutrality energy system.

## 1 Introduction

The global CO2 budget accounts for all natural[1] [2] [3-5] and human-driven CO2 sources and sinks within Earth's carbon cycle [5-10]. In the natural carbon cycle, carbon moves between the atmosphere, oceans, land, and living organisms[1] [2]. Photosynthesis absorbs CO2 from the air, storing it in plants and soils, while respiration, decomposition, and consumption by animals return it to the atmosphere. The natural carbon cycle also includes long-term carbon storage in soils and the lithosphere[3-5], which serves as a crucial mechanism for sustaining life on Earth and maintaining ecosystem functions [11].

However, human activities have significantly altered this natural carbon cycle. The combustion of fossil fuels, such as coal, oil, and gas, releases carbon stored in the lithosphere [9, 10, 12-14], while processes like deforestation reduce the land's capacity to absorb and store CO2 [11, 15-18]. This has resulted in the formation of an anthropogenic carbon cycle, where human-induced CO2 emissions disrupt the balance between carbon sources and sinks. Between 2010 and 2019, human activities released an average of $10.9 \pm 0.9$ petagrams of carbon (PgC) annually, with 46% of this CO2 absorbed by the atmosphere, 23% by the oceans, and 31% by terrestrial ecosystems. The remaining 0.1 PgC per year represents a budget imbalance within the uncertainties of the overall budget. Variations in the absorption rates of individual carbon sinks contribute to the complexities of the global carbon budget [19], but it is clear that anthropogenic emissions are the primary driver of the observed increase in atmospheric CO2 and the growing imbalance in the carbon cycle [20].

Human activities modify carbon flows and stocks through interactions with the land, ocean, and atmosphere [22, 23]. Carbon neutrality, a key benchmark for climate mitigation, is defined by the IPCC AR6 as achieving a balance between anthropogenic greenhouse gas emissions and removals. Current mitigation strategies, such as Nationally Determined Contributions (NDCs) and carbon neutrality targets set by countries, rely on emission accounting principles that focus on direct emissions (Scope 1), emissions from imported electricity (Scope 2) [24], and supply chain emissions (Scope 3) [25]. These measures, however, often overlook the role of replacing fossil fuels with carbon-neutral energy sources. In this context, the concept of "Scope 4 emissions" is gaining attention, referring to avoided emissions achieved by using carbon-neutral technologies or energy sources in place of fossil fuels, however, clear definition and calculation of Scope 4 emissions are still lacking, become challenges for implementing carbon mitigation and negative actions in both national and regional scales.

Here in this paper, we developed the new accounting framework for assessing the carbon-negative and carbon-neutral actions, through defining three key categories: **Carbon dioxide removal or negative carbon (C0)**, representing true carbon-negative processes; **carbon neutrality stock (C1)**, which refers to carbon captured in man-made carbon-containing products; and **carbon supply (C2)**. Given the fact that the final use of C1 products eventually re-releases captured carbon, we introduce the concept of **"carbon-neutral emissions" or "Scope 4 emissions"** associated with Recycling (CO2 emissions from the final use of carbon-containing products) or Replacement (the use of non-fossil-based materials and energy like renewables).

We estimated the calculation of neutralized carbon emissions or "Scope 4 emission" by global renewable electricity generation for the first time, and the results indicating the significant contribution by China, with total neutralized carbon emissions (2.15 Mt C/day) much higher than the U.S. (0.85 Mt C/day, 14.2%) and EU27 & UK (1.25 Mt C/day) together.

## 2 Methodology
### 2.1 Accounting framework for carbon negative and "Scope 4" Neutralized Carbon Emissions

There are four primary processes that enable carbon uptake through human activities: (1) photosynthesis in ecosystems, (2) natural weathering processes (such as the fossilization of limestone minerals), (3) chemical catalysis to produce carbon-containing compounds, and (4) direct $CO_2$ air capture and concentration via physical methods. The products of these processes are either carbon-containing compounds or, in rare cases, elemental carbon (e.g., diamond). Whether these products create a long-term carbon sink depends on their lifecycle: consumption or oxidation leads to carbon being released back into the atmosphere as $CO_2$, whereas products that avoid these fates and enter the lithosphere or biosphere can be classified as part of a "permanent" carbon sink or as a carbon-negative process. However, most carbon-containing products created through chemical processes (such as fuels or plastics) only temporarily store carbon. Ultimately, when these products are consumed or disposed of through burning, the captured carbon returns to the atmosphere.

By examining the carbon flows and stocks related to human activities (Fig. 1), it becomes clear that strategies for carbon uptake, carbon neutrality, and emission reduction can be pursued through various pathways. These include direct carbon capture from the atmosphere, enhancing natural carbon sinks, reducing emissions at the source, and optimizing carbon flows that indirectly improve carbon sinks and lower emissions. As a result, a straightforward framework for addressing anthropogenic carbon metabolism can be established by implementing these strategies in different contexts.

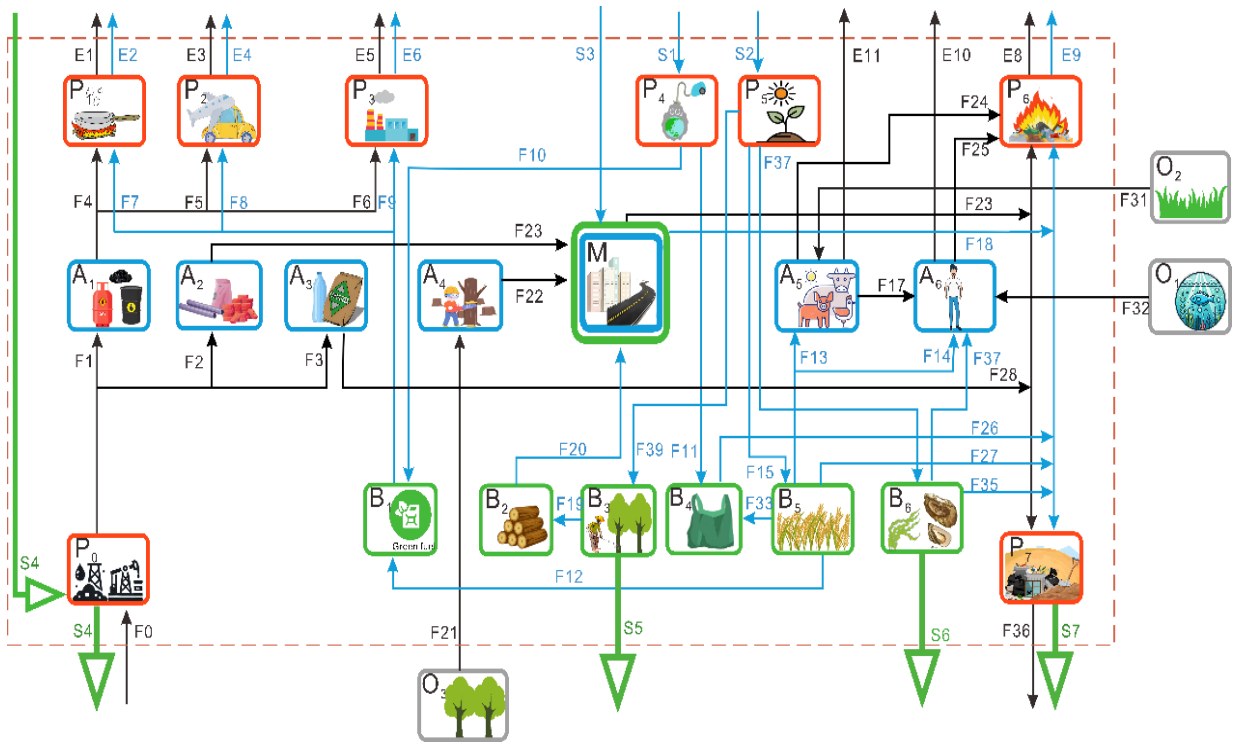

**Figure 1. Anthropogenic Carbon Flows.** B1–B6 represents C1 products. A1–A6 represents C2 products. P0-P7 indicate processes where carbon flows into or out of human society. Gray boxes represent external organic carbon pools (e.g., forests, soil). Black arrows indicate the movement of C2 carbon, which includes non-living and fossil fuel-based carbon. Blue arrows represent the movement of C1 carbon, which is associated with biologically derived products like biofuels and agriculture. Green arrows indicate the process of C0 carbon formation, such as photosynthesis.

B1: Biofuels and fuels produced from captured CO2;

B2: Timber products from artificial forests;

B3: Artificial forests;

B4: Products made from captured CO2;

B5: Agricultural crops;

B6: Marine farms (e.g., aquaculture);

A1: Fossil fuels;

A2: Non-fossil minerals and their derivatives;

A3: Fossil fuel-derived products;

A4: Timber products from natural forests;

A5: Livestock;

A6: Humans;

M (Buildings and roads): Represents infrastructure that contains embedded carbon, mainly from concrete and steel;

P0 (Mining): Refers to the extraction of minerals, contributing to carbon emissions;

P1 (Household fuel combustion): Carbon emissions from burning fuels for household energy;

P2 (Fuel combustion in the transportation sector): Emissions from fuel used in transportation;

P3 (Fuel combustion in factories): Industrial carbon emissions from fuel use;

P4 ($CO_2$ capture from the air): Processes involving the capture of $CO_2$ from the atmosphere;

P5 (Carbon absorption by plants via photosynthesis): Natural carbon sequestration through plant growth;

P6 (Waste incineration): Carbon emissions from burning waste materials;

P7 (Waste landfill): Long-term storage of carbon through the disposal of organic and synthetic waste;

O1 (Wild seafood): Carbon from marine ecosystems;

O2 (Natural pastures): Carbon stored in grazing lands;

O3 (Natural forests): Carbon stored in unmanaged forests;

To better understand and account for carbon-negative and carbon-neutral actions, we propose a new framework that defines three key categories: **Carbon dioxide removal or negative carbon (C0)**, representing true carbon-negative processes; **carbon neutrality stock (C1)**, which refers to carbon captured in man-made carbon-containing products; and **carbon supply (C2)**, which encompasses all types of man-made products. Importantly, C1 should be stable on a 100 year scale under standard temperature and pressure conditions (20°C, 1 atm). Under this definition, products like agricultural, animal husbandry, and fishery products fall under C2, while wood and wood products from plantation forests are classified as C1. C0 is a clear carbon sink, directly offsetting carbon emissions, while C1 may eventually become a permanent carbon sink or release $CO_2$, depending on how it is used or disposed of. The generation of C0 and C1 may also produce emissions due to the fossil energy or materials used in the process, necessitating a full life cycle assessment (LCA) that considers all carbon-related processes. Through such assessments, we can determine that a certain amount of carbon is initially taken up by anthropogenic activities. If the final emission is due to the combustion of C0 or C1 that was initially taken up by anthropogenic activities, such carbon emission can be considered "carbon neutrality emission" from a life-cycle perspective. We thus propose the concept of "Neutralized Emission" to encompass emissions offset by such carbon-neutral or carbon-negative processes, explained in the following

From a life cycle perspective, efforts to achieve carbon sinks or carbon-negative outcomes often involve increased energy use, which may generate additional emissions. However, emissions associated with the final use of C0 or C1 carbon-containing products can be considered carbon-neutral if those products are derived from human-induced carbon uptake processes (e.g., chemical, physical, photosynthesis, or enhanced weathering). Similarly, mitigation actions that avoid emissions by replacing emission-generating activities, such as using renewable energy instead of fossil fuels, have comparable effects. These actions do not directly result in additional emissions or carbon uptake but provide the necessary energy and materials for human activities without increasing the overall carbon emissions.

Based on this principle, we define two types of "**neutralized emissions**," also referred to as "**Scope 4 emissions**":

1) **Recycling**: Emissions associated with the final use (combustion or any process that release the carbon to the atmosphere as CO2) of carbon-containing products that are produced through human-induced carbon uptake processes, such as chemical catalysis, physical methods, photosynthesis, or enhanced weathering.

2) **Replacement**: The substitution of fossil fuels with materials and energy sources that do not rely on carbon-intensive processes, such as renewable energy.

These categories reflect a comprehensive approach to carbon neutrality and mitigation, focusing on recycling, emission reduction, and the replacement of fossil fuels to support sustainable human activities.

## 2.2 Estimates of global C1 and C2 products

We employed a socioeconomic approach to quantify the carbon stock in C1 and C2 products. This approach considers the material stocks within the socioeconomic system, including buildings, infrastructure, machinery, other products, livestock, and the human body. We first gathered inventory data for human and livestock species, material stocks in buildings, infrastructure, and machinery, as well as stored agricultural products.

The data for material stocks in buildings, infrastructure, and machinery were derived from an economy-wide dynamic material flow accounting model (MISO model v1). This model estimates the stock of materials such as asphalt, concrete, downcycled building minerals, primary sand and gravel, solid wood products, plastics, paper, steel, bricks, aluminum, copper, other metals, flat glass, and container glass. For bricks, aluminum, copper, and all other metals and glass, we assumed a carbon content of zero. To simplify carbon stock estimation, all aggregates and gravel from asphalt, concrete, downcycled materials, and primary sand and gravel were grouped together. Asphalt is composed of 95% aggregates and 5% bitumen, while concrete consists of 83% aggregates and 17% cement. Downcycled materials were initially divided into bricks, asphalt, and concrete, which were further broken down into aggregates, cement, and bitumen.

Agricultural crop stock data were sourced from the United States Department of Agriculture (USDA), covering 48 different crop products. Livestock data were sourced from the Food and Agriculture Organization (FAO), providing information on 16 livestock species. Finally, we calculated the carbon stock for each material by multiplying the material stock data by the carbon content of each material:

$$C_{i,y} = M_{i,y} \times a_i$$

where $C_{i,y}$ is the carbon stock of $i$th product in year y; $M_{i,y}$ is the material stock of $i$ in year $y$; $a_i$ is the carbon content ratio of $i$th material. In addition, the carbon content in carbonated cement can be quantified using a physico-chemical model that accounts for the thickness of the cement materials, exposure conditions of different strength categories, and atmospheric $CO_2$ concentrations across various geographic regions.

## 2.3 Calculation of "scope 4 emissions"

We collected over 2 million records related to power generation, encompassing various primary energy types (e.g., coal, gas, oil, solar, hydro) across national to provincial spatial scales and from monthly to sub-hourly temporal scales. Due to the prevalence of missing values in the raw data, we applied data cleaning operations. Outliers, defined as values above the 75th percentile plus 1.5 times the interquartile range (IQR) or below the 25th percentile minus 1.5 IQR, were replaced [21]. We then addressed missing values, including the replaced outliers, through linear interpolation. For detailed information on the data processing, please refer to our previous research [21].

We calculated the daily neutralized emissions for different countries. We began by aggregating the total electricity generation from all non-fossil fuel types for each day, denoted as G(R). Assuming that G(R) would have been generated by thermal power plants instead, the corresponding neutralized carbon emissions, E(AE), can be determined. By considering the daily electricity generation ratio (percentage) of coal, oil, and gas, represented by P(coal), P(oil), and P(gas), along with their respective emission factors, EF(coal), EF(oil), and EF(gas), we calculated the neutralized carbon emissions, E(AE), as follows:

$$E(AE) = G(R) * [ P(coal) * EF(coal) + P(oil) * EF(oil) + P(gas) * EF(gas) ]$$

## 3. Results
### 3.1 Global C0, C1 and C2

We estimated the C0, C1 and C2 at global scale, it's clearly that C0 and C1 are very limited when comparing the C2 in global scale, indicating the urgent needs for further development of carbon neutrality and carbon negative.

**C0:** Key methods include afforestation, enhanced weathering, cement carbonation, and direct air capture (DAC). Enhanced weathering, by applying crushed silicate rocks to land, sequesters $CO_2$ while improving soil and crop yields. DAC for $CO_2$-enhanced oil recovery can sequester 35 million tons of $CO_2$ annually, with a global potential of 2 billion tons per year. To reach carbon neutrality within a century, C0 must increase 100 times.

**C1:** C1 includes biofuels, plantation forests, and carbonated cement but excludes fossil fuels and food. Plantation forests, though susceptible to wildfires, are important for wood and paper production. Biomass energy is vital for high-energy industries like aviation. Cement carbonation sequesters $CO_2$ by reacting with alkaline substances. In 2015, carbon stock was 2.9 Pg C for weathered cement, 0.47 Pg C for biomass products, 12.9 Pg C for artificial forests, and 0.16 Pg C for crop products.

**C2:** C2 includes carbon from all manufactured materials, both C1 (e.g., biofuels, carbonated cement) and non-C1 (e.g., steel, fossil fuels). Infrastructure, such as roads and buildings, contains both C1 and C2 products. Based on methods provided in relevant literature, we calculated the carbon content of four typical C2 products (Figure 2): non-fossil minerals, fossil fuels and their derivatives, carbonated cement, and solid wood products. The total carbon content increased 17-fold from 1900 to 2015, rising from 2.5 PgC to 42 PgC. Non-fossil mineral C2 stocks showed the largest increase, growing from 0.6 PgC in 1900 to 25.2 PgC in 2015, with its share of total C2 rising from 23.9% to 59.9%. From 1900 to 2015, the C2 stock of solid wood products increased from 1.8 PgC to 6.9 PgC, although its proportion of total C2 dropped from 71.9% to 16.5%. In 2015, fossil fuels and their derivatives accounted for 6.4 PgC, making up 15.0% of total C2. Carbonated cement had a carbon stock of 2.7 PgC, representing 6.5% of the total C2 storage.

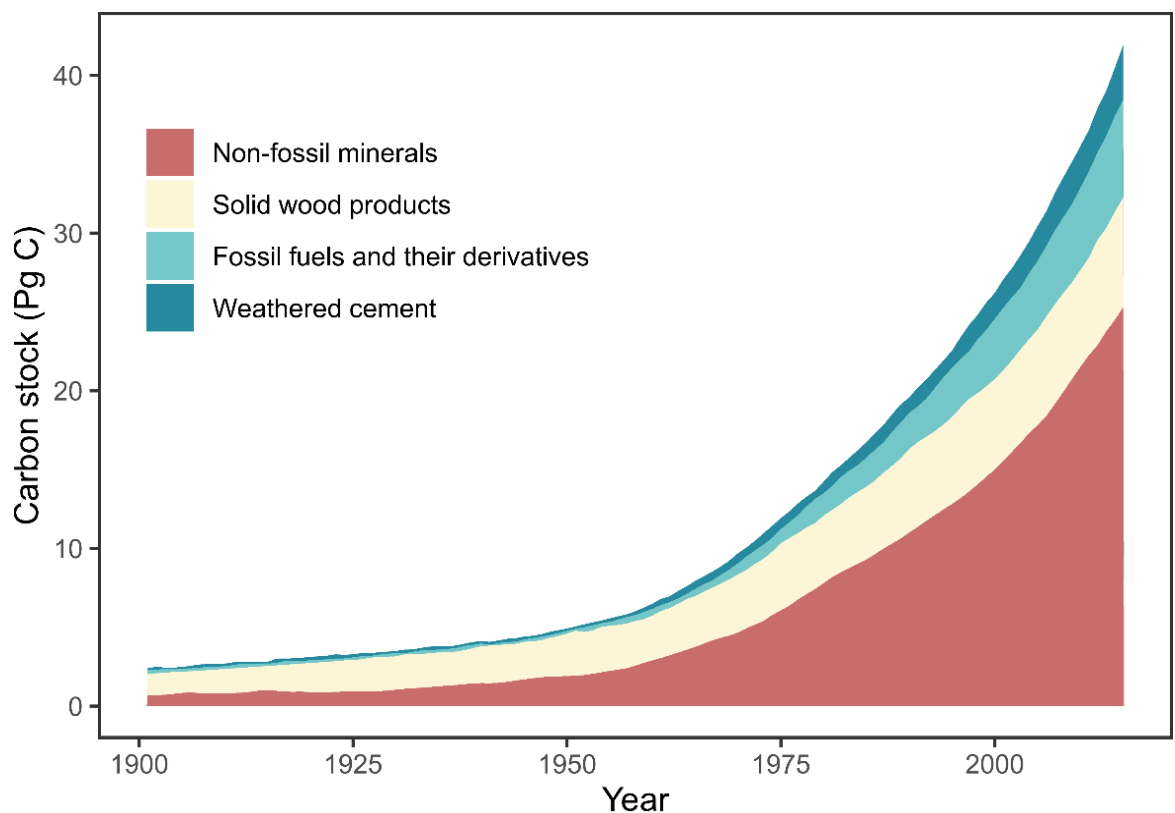

**Figure 2.** Global C2 products from 1900 to 2015

**Table 1.** Main C0 processes and their annual carbon sequestration amount.

| Process | Classification | Definition | Carbon storage per year (Pg C a$^{-1}$) |
|---|---|---|---|
| Physical process | C0 | Artificially captured CO2 for enhanced oil recovery and geological sequestration | 0.035 |
| Chemical process | C0 | Landfilled chemicals produced from air-captured CO2 | Not determined |
| Biomass product | C0 | Buried biochar from afforestation and CO2 sequestered by shellfish in aquaculture | Not determined |
| Mineral weathering | C0 | Landfilled weathered cement | Not determined |

**Table 2.** Anthropogenic Carbon Products.

| Products | Classification | Definition | Carbon stock in 2015 (Pg C) |
|---|---|---|---|
| Weathered cement | C1 | Cement in concrete which is undergoing CO2 weathering | 2.9 |
| Biofuel | C1 | Bioenergy produced from carbon captured through plant photosynthesis | Not determined |
| Biomas product | C1 | Products derived from plant fibers, such as paper | 0.47 |
| Chemical from CO2 | C1 | Chemicals produced using CO2 captured from air or flue gas | Not determined |
| Fuel from CO2 | C1 | Fuels produced using CO2 captured from air or flue gas | Not determined |
| Artifical forest | C1 | Carbon content of forests formed through afforestation and reforestation | 12.9 |
| Wood product from artificial forest | C1 | Various solid wood products from plantation timber | X1 |
| Agriculture crop products | C1 | Agricultural crop products such as wheat, rice, etc | 0.16 |
| Fossil fuels | C2 | Fuels directly used for providing energy or products made from fossil fuels | 6.4 |
| Wood product from natural forest | C2 | Various solid wood products made from natural forest timber | X2 |
| Non-fossil minerals | C2 | Non-fossil minerals such as iron/steel and aggregates | 25 |
| Human body | C2 | All living humans | 0.060 |
| Livestock | C2 | Livestock such as cattle, sheep, chickens, and ducks | 0.029 |

Note that X1 + X2 = 6.9 PgC.

## 3.2 Global Scope 4 neutralized emissions by electricity power generation

The "neutrality emissions" or "Scope 4 emissions," associated with Recycling ($CO_2$ emissions from the final use of carbon-containing products), Reduction (decreased carbon intensive activities), and Replacement (the use of non-fossil-based materials and energy like renewables). Here we show the case of Replacement, that we calculated the global Carbon Neutral Emissions (Scope 4 emissions) through using non-fossil energy sources in place of fossil fuels.

Global Scope 4 neutralized emissions have seen significant growth from 2019 to 2024 (Fig. 3 and 4), with total neutralized $CO_2$ emissions reaching 11923.87 Mt C during this timeframe. Nevertheless, these emissions have not yet exceeded the total direct $CO_2$ emissions resulting from fossil fuels (coal, gas, and oil), which were 16366.07 Mt C for the same period, averaging 8.24 Mt C/day. Despite this, advancements in neutralized emissions signify a critical milestone in global climate mitigation initiatives.

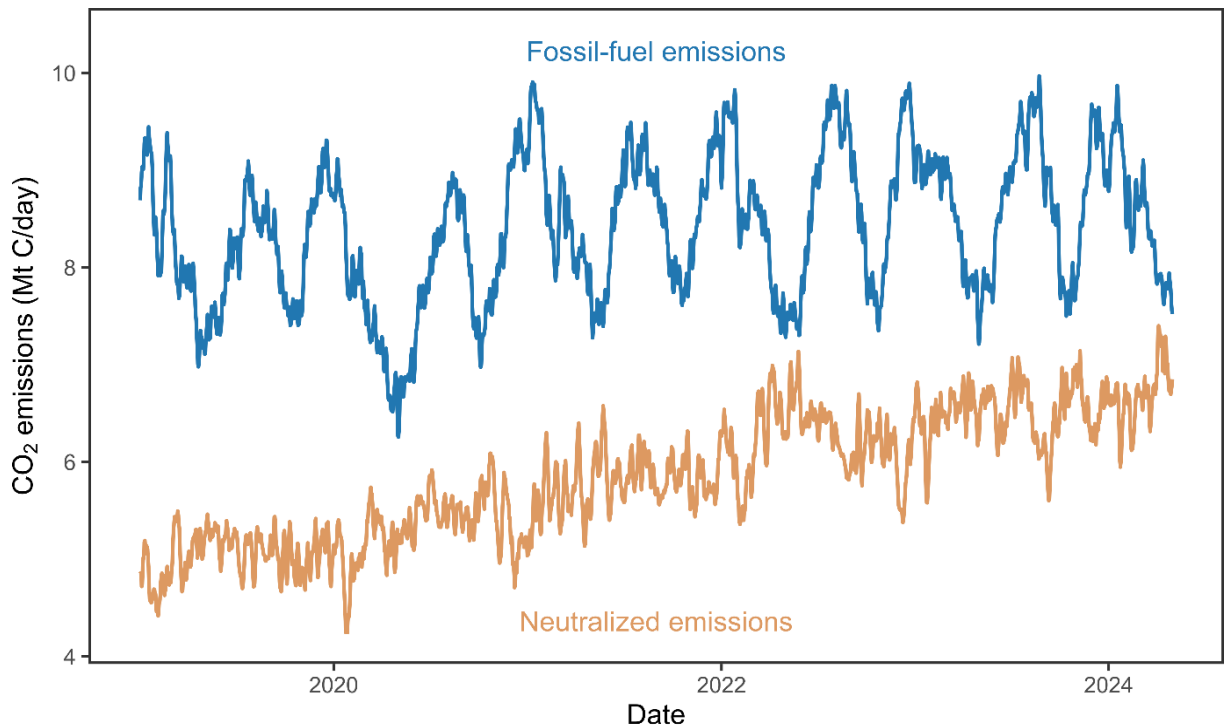

**Figure 3. Global Neutralized Carbon Emissions due to renewable and nuclear power generation.** renewable and nuclear power generation is represented in red, while the time series of global fossil fuel emissions is shown in black.

The calculation of scope 4 neutralized emissions indicated significant contribution to climate mitigation from China (Figure 4). Among various countries, China stands out as the largest contributor to global neutralized CO2 emissions, averaging 2.15 Mt C/day and accounting for an impressive 36.7% of the total. This places China significantly ahead of the EU27 & UK, which contribute 1.25 Mt C/day (21.3%). Following behind are the United States (0.85 Mt C/day, 14.2%), Brazil (0.44 Mt C/day, 7.4%), and Russia (0.38 Mt C/day, 6.5%), with the remainder of the world (ROW) contributing 0.35 Mt C/day (6.3%).

The average annual growth rate (AGR) of global neutralized CO2 emissions is 3.62%, outpacing the 1.49% AGR of fossil CO2 emissions. This highlights significant progress, particularly driven by China's investments in reducing fossil fuel dependency. However, the growth in neutralized CO2 emissions has slowed. After peaking at 6.5% between 2020 and 2021, the increase dropped to 3.2% in early 2024 compared to 2023. This deceleration is linked to reduced power demand, which peaked at 7.2% in 2021 but fell to 1.6% in early 2024. Factors such as economic inflation post-pandemic and the Russia-Ukraine conflict have contributed to this reduced demand, limiting the rise in neutralized CO2 emissions.

China's contribution remains pivotal in global efforts to curb emissions, underscoring the nation's role in the ongoing energy transition.

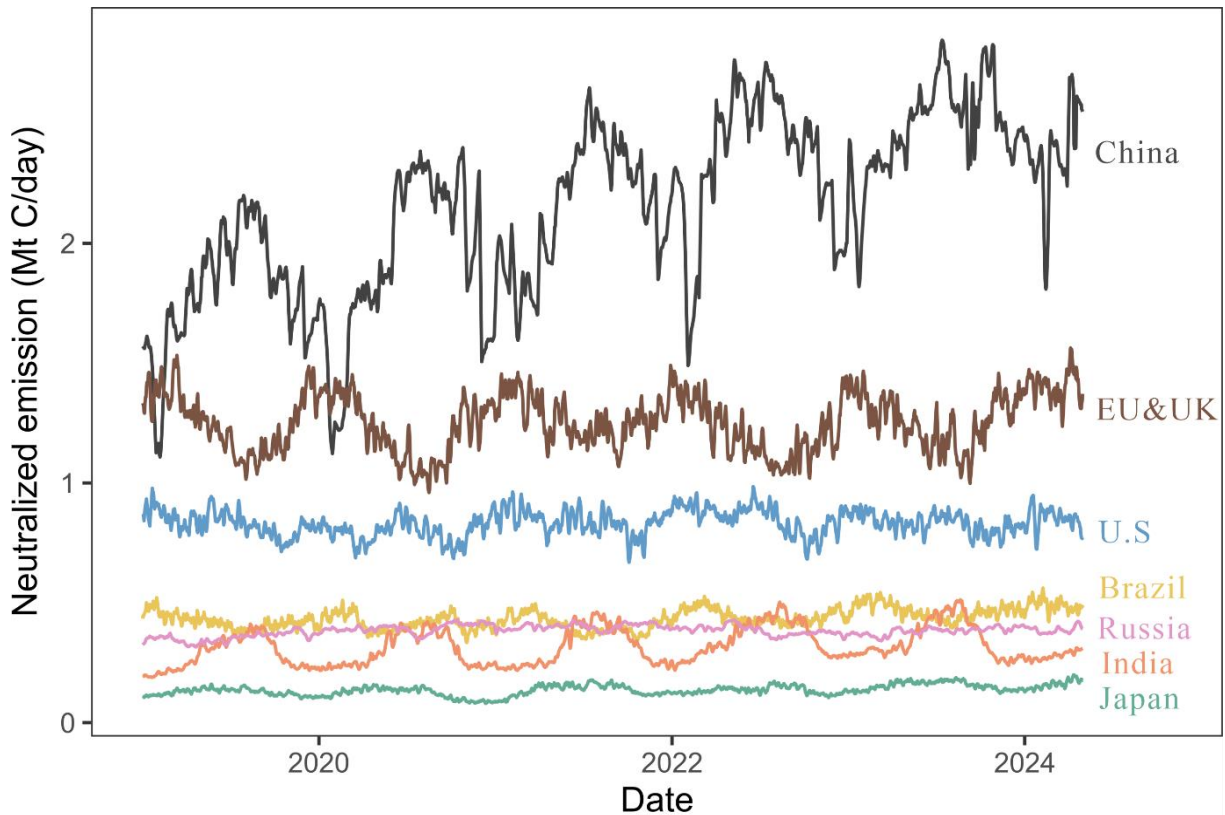

**Figure 4. National Neutralized Carbon Emissions due to renewable and nuclear power generation.** U.S.: United States; EU&UK: European Union and the United Kingdom.

## 4. Discussion and conclusions

We propose a framework to account for carbon-negative and carbon-neutral actions by introducing key concepts: Carbon Negative (C0), Carbon Neutrality Stock (C1), which represents the carbon stored in man-made products derived from air carbon capture, and Carbon Supply (C2), the total carbon embedded in all man-made products. Since the ultimate use of C1 products releases the captured carbon back into the atmosphere, we define "carbon-neutral emissions" or "Scope 4 emissions" to encompass emissions from Recycling (CO2 emissions from the final use of carbon-containing products) and Replacement (the substitution of fossil-based materials and energy with renewable alternatives). This framework promotes carbon mitigation efforts and contributes to sustainable carbon neutrality pathways.

A critical component of this framework is the contribution of China, which plays a leading role in global neutralized CO2 emissions. China currently contributes 36.7% of neutralized CO2 emissions, averaging 2.15 Mt C/day, making it the largest contributor. This outpaces the EU27 & UK (21.3%) and other major economies, such as the United States (14.2%). China's significant contribution underscores its pivotal role in carbon mitigation through advancements in renewable energy and carbon capture technologies. This contribution highlights China's importance in closing the carbon mitigation gap and achieving long-term global carbon neutrality.

Such a method encourages mitigation actions across production and consumption sectors, including industry, transportation, and buildings, all crucial for carbon reduction. Studies on achieving deep decarbonization in these sectors highlight the need for technological adoption, behavioral changes, and policy measures that improve energy efficiency and control demand. By clearly distinguishing between Carbon Negative (C0), Carbon Neutrality Stock (C1), and Carbon Supply (C2), this framework provides a structured approach to categorize carbon flows in man-made products. This enhances transparency in tracking emissions and clarifies the role of different products in the carbon cycle.

The concept of "Scope 4 emissions" addresses a gap in existing carbon accounting by focusing on emissions from recycling, reduction, and replacement processes, areas often overlooked in traditional frameworks. By accounting for the carbon-neutral potential of recycled and reduced emissions, the framework encourages businesses and industries to adopt circular economy practices, driving innovation in carbon capture and recycling technologies. It also promotes the substitution of fossil fuels with renewable energy and reductions in carbon-intensive activities, encouraging organizations to quantify their contributions toward carbon neutrality.

Finally, linking carbon-negative and carbon-neutral activities to measurable outcomes helps develop practical strategies for achieving carbon neutrality. This supports better policy formulation and corporate strategies that promote sustainable practices across various sectors, aiding global efforts to limit climate change.

Given the limited amount of C0, increasing C0 through natural, technological, and hybrid methods is essential. Natural approaches include afforestation, soil carbon sequestration, and wetland restoration, while technological methods, such as direct air capture (DAC) and enhanced weathering, remove CO2 from the atmosphere. Hybrid methods, like agroforestry and biochar application, offer combined strategies. Additionally, ocean-based techniques, such as seaweed cultivation and alkalinity enhancement, contribute to carbon sequestration. China's leadership in deploying these approaches emphasizes its critical role in advancing global carbon-neutral and carbon-negative strategies.

**Declarations**

The authors declare that they have no conflict of interest.

**Author contribution statement**

Z.L. designed and wrote the paper, G.Q.W. developed the concept of C0, C1 and C2. All the authors have approved the final manuscript.

**Data availability**

The data that support the findings of this study could be downloaded from https://power.carbonmonitor.org/